\documentclass[12pt]{article} 
\usepackage{graphicx}
\newcommand{\be}{\begin{equation}}
\newcommand{\ee}{\end{equation}}
\newcommand{\ba}{\begin{array}}
\newcommand{\ea}{\end{array}}

\newcommand{\grad}{{\rm \; grad\;}}

\newcommand{\bea}{\begin{eqnarray*}}
\newcommand{\eea}{\end{eqnarray*}}
\newcommand{\bean}{\begin{eqnarray}}
\newcommand{\eean}{\end{eqnarray}}
\newcommand{\proof}{\vspace{1ex}\noindent{\em Proof}. \ }
\def\ds{\displaystyle}
\def\nm{\noalign{\medskip}}

\newtheorem{lemma}{Lemma}[section]

\newtheorem{theorem}{Theorem}[section]
\newtheorem{corollary}{Corollary}[section]
\newtheorem{proposition}{Proposition}[section]
\usepackage{amsmath, amssymb}
\usepackage{latexsym}
\newcommand{\R}{\mathbb{R}}

\def\Box{\leavevmode\vbox{\hrule
     \hbox{\vrule\kern5pt\vbox{\kern5pt}%
           \vrule}\hrule}}
\renewcommand{\square}{\hfill$\Box$}
\begin{document}
\title{Asymptotic behaviors for eigenvalues and eigenfunctions associated to Stokes operator
 in the presence of small boundary perturbations}

\author{Christian Daveau \thanks{
 Department of Mathematics, CNRS (UMR 8088),
 University of Cergy-Pontoise,
 2 avenue Adolphe Chauvin,
 95302 Cergy-Pontoise Cedex, France. (Email: christian.daveau@u-cergy.fr)}
\and Abdessatar Khelifi \thanks{
D\'epartement de Math\'ematiques, Universit\'e des Sciences de
Carthage, Bizerte, Tunisie. (Email: abdessatar.khelifi@fsb.rnu.tn)}}

\maketitle \abstract{We consider the Stokes eigenvalue problem in a
bounded domain of $\mathbb{R}^3$ with Dirichlet boundary conditions.
The aim of this paper is to advance the development of high-order
terms in the asymptotic expansions of the boundary perturbations of
eigenvalues, eigenfunctions and eigenpressures for the Stokes
operator caused by small perturbations of the boundary. Our
derivation is
rigorous and proved by layer potential techniques.}\\



\section{Introduction}\label{prob-form}\label{section1}
The field of eigenvalue problems under shape perturbation has been
an active research area for several decades. Several related
problems belong to Stokes systems, which are further subdivided by
assumptions on the underlying media and on the Dirichlet boundary
conditions. The main objective of this paper is to present a
schematic way to derive high-order asymptotic expansions for both
eigenvalues and eigenfunctions for the Stokes operator caused by
small perturbations of the boundary. The properties of eigenvalue
problems under shape deformation have been a subject of
comprehensive studies \cite{albert1,albert2,babuska,ortega1} and the
area continues to carry great importance
\cite{james,kohr,medkova,jia,khelifi}. A substantial portion of these
investigations discusses the properties of smoothness
and analyticity of eigenvalues and eigenfunctions with respect to perturbations.\\

 Let $\Omega\subset\mathbb{R}^3$ be a bounded open domain with
boundary of class $C^2$. We consider the following eigenvalue
problem for the Stokes system with homogeneous boundary conditions:

\begin{equation}\label{system1}
 \begin{cases}
   -\Delta v+\nabla.p=\lambda v & \mbox{in }\Omega\\
 \nabla.v=0 & \mbox{in }\Omega \\
  v=0 & \mbox{in }\partial\Omega. \\
  \end{cases}
\end{equation}
Here $v = (v_1, v_2,v_3)$ denotes the velocity field, while the
scalar function $p$ is the pressure.\\
 It is well known that this eigenvalue problem admits a sequence
of a no decreasing positive eigenvalues $0 < \lambda_1 \leq
\lambda_2 \leq \cdots\leq \lambda_n\leq\cdots$ tending to infinity
as
$n\to+\infty$.\\
The eigenfunctions $\{v_n\}_{n\geq1}\subset(H_{0}^{1}(\Omega))^2$
and the eigenpressures $\{p_n\}_{n\geq1}\subset L^2(\Omega)$ may be
taken so that $\{v_n\}_{n\geq1}$ constitutes an orthonormal basis of
$$H(\Omega):=\{v\in (L^2(\Omega))^3:\quad \nabla\cdot v=0 \mbox{ in
}\Omega,\mbox{ and } u=0 \mbox{ on } \partial\Omega\}.$$ The
pressure
$p$ is determined up to an additive constant.\\
We assume that $\Omega$ has a small and smooth deformation and that
the boundary of the deformed domain $\Omega_\delta$ is the set of
points $\tilde{ x}$ defined by
\begin{equation}\label{parametr}
\partial {\Omega}_\delta:=\{\tilde{ x}=x+\delta\rho(x)\nu(x),\quad x\in\partial\Omega\}
\end{equation}
where $\nu(x)$  is the outward normal vector on on $\partial\Omega$
and $\rho(x)$ is a real function in $C^2(\partial\Omega)$ that
satisfies
\begin{equation} \label{est-rho}
\|\rho(x)\|_{C^2(\partial\Omega)} <1.
\end{equation}
Obviously, the domain $\Omega_{\delta}$ is of class $C^2$ and the
Dirichlet eigenvalue problem for the Stokes system can be defined in
$\Omega_{\delta}$ as
well.\\

 In this paper, we derive the asymptotic of eigenvalues, eigenfunctions and the eigenpressures solutions to the Stokes system:
\begin{equation}\label{system2}
 \begin{cases}
   -\Delta v_\delta+\nabla.p_\delta=\lambda_\delta v_\delta & \mbox{in }\Omega_\delta\\
 \nabla.v_\delta=0 & \mbox{in }\Omega_\delta \\
  v_\delta=0 & \mbox{in }\partial\Omega_\delta.
  \end{cases}
\end{equation}
Here we suppose that the eigenvalue $\lambda_0$ is simple. Then the
eigenvalue $\lambda_\delta$ is simple and is near to $\lambda_0$
associated to the normalized eigenfunction $v_\delta$.\\
 To the best of our
knowledge, this is the first work to rigorously investigate Stokes
eigenvalue problem in the presence of the perturbation
(\ref{parametr}) and derive high-order terms in the asymptotic
expansion of $\lambda_\delta-\lambda_0$ and $v_\delta-v_0$
 when $\delta\to 0$. However, by the same method, one can derive asymptotic
formula for the Neumann problem as well.\\ Zuazua and Ortega have
proved in \cite{ortega1} the regularity of the eigenvalues and
eigenfunctions of the Stokes system with respect to the perturbation
parameter, by using the Lyapunov-Schmidt method. Their proofs are
essentially inspired in the work of J. Albert \cite{albert1,albert2}
for the Laplace operator. Our analysis and uniform asymptotic
formulas of eigenvalues and eigenfunctions, which are represented by
the single-layer potential involving the Green function, are
considerably different from those in \cite{james,ortega1}.

This paper is organized as follows. In section 1, we describe the
main problem in this work. In section 2, we develop a boundary
integral formulation for solving the eigenvalue problem
(\ref{system2}), and we present some preliminary results. In the
last section, we derive by layer potentials techniques formal
asymptotic expansion for both eigenvalues and eigenfunctions of the
Stokes operator.
\section{Integral equations method}
We now develop a boundary integral formulation for solving the
eigenvalue problems (\ref{system1}) and (\ref{system2}). The
components of the fundamental Stokes tensor $\Gamma$ and those of
the associated pressure vector $P$, which determine the fundamental
solution $(\Gamma, P)$ of the Stokes system in $\mathbb{R}^3$, are
given by (see for instance \cite{ladyzh})
\begin{equation}\label{system3}
 \begin{cases}
  \Gamma_{ij}(\lambda,x)=\frac{1}{4\pi}\frac{\delta_{ij}e^{i\lambda|x|}}{|x|}+
  \frac{1}{4\pi\lambda^2}\partial_{x_i}\partial_{x_j}(\frac{e^{i\lambda|x|}-1}{|x|^3})\\
P_{i}(x)= \frac{1}{4\pi}\frac{x_i}{|x|^3}.
  \end{cases}
\end{equation}
We recall that the ith row, $\Gamma_i$ of $\Gamma$ satisfies
\begin{equation}\label{system4}
 \begin{cases}
   -\Delta \Gamma_i+\nabla.P_i(x)-\lambda\Gamma_i=e^{i}\delta(x) & \mbox{in }\mathbb{R}^3\\
 \nabla.\Gamma_i=0 & \mbox{in }\mathbb{R}^3 ;
  \end{cases}
\end{equation}
where ($e^i; i = 1, 2, 3$) is the orthonormal basis of
$\mathbb{R}^3$.
\subsection{The potential theory for the Stokes system}
Let us denote by $\varphi= ( \varphi_1, \varphi_2,\varphi_3)$ a complex vector-valued function in the class $C^0(\partial\Omega)$.\\

 The hydrodynamic single layer potential with density
$\varphi\in C^0(\partial\Omega)^3$ is the vector function
$\mathcal{S}(\lambda)\varphi(x)$ defined by
\begin{equation}\label{single}
\mathcal{S}(\lambda)\varphi(x):=\int_{\partial\Omega}\Gamma(\lambda,|x-y|)\varphi(y)~d\sigma(y),\quad
x\in\mathbb{R}^3\backslash\partial\Omega.
\end{equation}

The pressure term $\mathcal{Q}$ corresponding to the single layer
potential is the function given by
$$
\mathcal{Q}\varphi(x):=\int_{\partial\Omega}P(x,y)\varphi(y)~d\sigma(y),\quad
x\in\mathbb{R}^3\backslash\partial\Omega.
$$
For a cureful stady of these potentials, one can refer to
\cite{ladyzh}, \cite{warn}, \cite{kohr}.

Taking into account the well known properties of Green function
$\Gamma$, one obtains the result that the pair
($\mathcal{S}\varphi,\mathcal{Q}\varphi$) are smooth functions in
each of the domains $\Omega$ and
$\mathbb{R}^3\backslash\overline{\Omega}$ respectively.
Also these functions are classical solution to the Stokes system (\ref{system1}).\\

The continuity and jump relations of the Stokes surface potentials
on the boundary $\partial\Omega$ are described in the following
proposition (see \cite{nicolas} pp. 41-42 or  \cite{warn} p. 66 ):
\begin{proposition}\label{prop-saut}
Let $\varphi\in C^0(\partial\Omega)^3$ and let $\mathcal{S}$ denotes
the surface potential defined in (\ref{single}). Then on the
boundary $\partial\Omega$ the following continuity and jump
relations are satisfied:
$$
\begin{array}{c}
 \ds (\mathcal{S}(\lambda)\varphi)\big|_+= (\mathcal{S}(\lambda)\varphi)|_-=  \mathcal{S}(\lambda)\varphi \\
\ds
\frac{\partial\mathcal{S}(\lambda)(\varphi)}{\partial\nu}(x)\big|_{\pm}=
\pm\frac{\varphi(x)}{2}+\int_{\partial\Omega}\frac{\partial\Gamma(\lambda,|x-y|)}{\partial\nu(x)}\varphi(y)~d\sigma(y).
\end{array}
$$
\end{proposition}
\subsection{The boundary integral formulation of the problem} In this section, we
give a boundary integral formulation in order to solve the
eigenvalue problems  (\ref{system1}) and (\ref{system2}).
\begin{proposition}\label{diffeo} Suppose $\rho$ satisfies  (\ref{est-rho}).
Then, there exists $\delta_0 >0$ such that  the map
$\Psi_{\delta}(x)$ defined by
\begin{eqnarray} \label{psi-d}
\Psi_{\delta}(x) = x + \delta \rho(x)\nu(x)
\end{eqnarray}
is a $C^2$- diffeomorphism from $\partial \Omega$ to $\partial
\Omega_{\delta}$ for $\delta < \delta _0$. In addition, the
following equality  holds
\begin{eqnarray}\label{jacobien}
\det \left( \nabla \Psi_\delta \right)  &= & 1 + \rm{tr} \nabla(\rho
\nu)\delta + \frac{1}{2}\big[\big(\rm{tr}
\nabla(\rho\nu)\big)^2-\rm{tr} \big(   \nabla (\rho \nu)^2
\big)\big] \delta^2+\det\left(  \nabla(\rho \nu)  \right) \delta^3,
\end{eqnarray}
where $ \rm{tr}$ means the trace of a matrix. Moreover, we have  $
\rm{tr}  \nabla(\rho \nu)= \rm{div}(\rho\nu)$.
\end{proposition}
\proof Recall that $\partial \Omega  $ is $C^3$ and so $\nu(x)$ is a
$C^2$ vector-valued function. Since the function $\rho(x)$ is $C^2$,
the map   $\Psi_{\delta}(x)$ is also $C^2$. A simple calculation
yields the equality~(\ref{jacobien}).  Consequently, for $\delta$
small enough the map $\Psi_{\delta}(x)$ is a $C^2$- diffeomorphism
from $\partial \Omega$ to $\partial \Omega_{\delta}$.
\square\\

We further denote $ \Psi_{\delta}^{-1}$ the reciproque function of
$\Psi_{\delta}(x)$. Thanks to $ \Psi_{\delta}^{-1}$, we can define
the operator $ \mathcal{A}_\delta(\lambda)$ as follows:
\begin{equation}\label{op1}
 \mathcal{A}_\delta(\lambda)\varphi(x)=\big(\mathcal{S}_{\Omega_\delta}(\lambda)\varphi(\Psi_{\delta}^{-1})\big)(\Psi_{\delta}(x)),\quad \varphi\in (L^2(\partial\Omega))^3
\end{equation}
where $\mathcal{S}_{\Omega_\delta}(\lambda)$ is the hydrodynamic single layer potential given by (\ref{single}) when we have replaced the boundary $\partial\Omega$ by $\partial\Omega_\delta$.\\

For $i,j\in\{1,2,3\}$, we can define the $j^{th}$-component of the
vector-valued function $ \mathcal{A}_\delta(\lambda)$ as follows:
\begin{equation}\label{OP1*}
 (\mathcal{A}^\delta(\lambda)\varphi)_{j}(x)=\Big(\big(\mathcal{S}_{\Omega_\delta}
(\lambda)\varphi\big)_{j}(\Psi_{\delta}^{-1})\Big)(\Psi_{\delta}(x)),\quad
\varphi\in (L^2(\partial\Omega))^3.
\end{equation}
The $j^{th}$-component of the single layer potential
$\mathcal{S}_{\Omega_\delta} (\lambda)$ is given by
\begin{equation}\label{single-delta}
\big(\mathcal{S}_{\Omega_\delta}(\lambda)\varphi\big)_{j}(\tilde
x):=\int_{\partial\Omega_\delta}\Gamma_{ji}(\lambda,|\tilde x-\tilde
y|)\varphi_i(\tilde y)~d\sigma_{\delta}(\tilde y),\quad \tilde
x\in\mathbb{R}^3\backslash\partial\Omega_\delta,j=1,2,3,
\end{equation}
where $\varphi_i$ is the $i^{th}$-component of the vector-valued
function $\varphi$. Using Proposition \ref{diffeo}, relations
(\ref{OP1*})-(\ref{single-delta}) and the continuity relations given
by Proposition \ref{prop-saut}, we obtain for $x\in\partial\Omega$
that
\begin{equation}\label{op1*}
 \big(\mathcal{A}^\delta(\lambda)\varphi\big)_{j}(x)=\int_{\partial\Omega}\Gamma_{ji}(\lambda,|\Psi_{\delta}( x)-\Psi_{\delta}( y)|)\det \left( \nabla \Psi_\delta(y) \right)\varphi_i( y)~d\sigma( y),\quad
j=1,2,3.
\end{equation}

Let $\mathcal A_0$ the operator defined as in (\ref{op1}) by $$
 \mathcal{A}_0\phi=\mathcal{S}(\lambda)\phi,$$ where $\phi\in (L^2(\partial\Omega))^3$. Then, we have the following result,
  which is a slight variation of the Lemma 6.1 due to Ammari and Triki \cite{AT} on the scalar eigenvalue problem for the
Laplacian.
\begin{proposition}\label{prop1} The
operator-valued function $\mathcal{A}_0(\lambda):
H^{-1/2}(\partial\Omega)^3\to H^{1/2}(\partial\Omega)^3$ is Fredholm
of index zero in $\mathbb{C}\backslash i\mathbb{R}^-$. In addition
the Dirichlet eigenvalues of the Stokes system (\ref{system1}) are
exactly its real zeros.
\end{proposition}

From Proposition \ref{prop1} we know that if $\lambda_0$ is an
eigenvalue of (\ref{system1}) then $\lambda_{0}$ is a real zero of
$\mathcal{A}_{0}(\lambda)$. Moreover, for $\epsilon_0$ small enough,
the function $\mathcal{A}_{0}^{-1}(\lambda)$ is meromorphic in
$\overline{D_{\epsilon_{0}}} (\lambda_{0})$, where
$D_{\epsilon_{0}}(\lambda_{0})$ the disc of center $\lambda_{0}$ and
radius $\epsilon_{0} $,  and $\lambda_{0}$ is its unique pole in
$\overline{D_{\epsilon_{0}}}$. Furthermore we have the Laurent
expansion \bean \label{eq10} \mathcal{A}_{0}^{-1}(\lambda) =
(\lambda-\lambda_{0})^{-1}\ell_{0} + R_{0}(\lambda), \eean where $
\ell_{0}: Ker\mathcal{A}_{0}(\lambda_{0}) \to Ker
\mathcal{A}_{0}(\lambda_{0}) $, and $R_{0}(\lambda)$ is a
holomorphic function.\\

Our main results in this section are summarized in the following
theorem.
\begin{theorem}\label{op-lauren1}
Suppose that the eigenvalue $\lambda_0$ of (\ref{system1}) is with
multiplicity 1. Then, there exists a positive constant
$\delta_{0}(\epsilon_{0})$ such that for
 $|\delta| < \delta_{0}$, the operator-valued function $\lambda \mapsto \mathcal{A}_{\delta}(\lambda)$ has
a real zero $\lambda(\delta)$ in
$\overline{D_{\epsilon_{0}}}(\lambda_{0})$. This zero is exactly the
eigenvalue of the perturbed eigenvalue problem (\ref{system2}), and
is an analytic function with respect to $\delta$  in
$]-\delta_{0},\delta_{0}[ $. It satisfies
 $\lambda(0) = \lambda_{0}$. Moreover, the following assertions hold: \bean \label{eq9}
\begin{array}{llll}
   \ds  \mathcal{A}_{\delta}^{-1}(\lambda) =
(\lambda - \lambda(\delta))^{-1}\ell(\delta) + R_{\delta}(\lambda),\\
\nm
 \ds  \ell(\delta):  Ker(\mathcal{A}_{\delta}(\lambda(\delta))
\to Ker(\mathcal{A}_{\delta}(\lambda(\delta)),
\end{array}
\eean where $R_{\delta}(\lambda)$ is a holomorphic function with
respect to
 $(\delta,\lambda) \in ]-\delta_{0},\delta_{0}[ \times D_{\epsilon_{0}}(\lambda_{0})$.
\end{theorem}

\section{Asymptotic behaviors}
\subsection{High-order terms in the expansion of $\mathcal{A}_{\delta}$}
We now present some basic results related to shape perturbation. To
begin,  let $ (\tau_1(x),\tau_2(x))$ be the orthornormal basis of
the tangent plan to the surface $\partial  \Omega$ at  a regular
point  $x $.   Their
 cross product is so orthogonal to  $\partial  \Omega$ at  the point $x$.  By
changing their order,  we can assume that $\tau_1\times \tau_2$ is a
vector pointing  towards the exterior of the surface  $\partial
\Omega$. Then dividing it
 by its length yields the unit normal vector $\nu(x)$, that is
\begin{equation}\label{normal0}
\nu_0(x)=\frac{\tau_1(x)\times \tau_2(x)}{|\tau_1(x)\times
\tau_2(x)|},
\end{equation}
for $x\in \partial \Omega$.
Evidently $\nu_0=\nu$, where $\nu$ was introduced in section 1.\\

 Set $$\tau_1^\delta = \grad \Psi_\delta\cdot \tau_1,\quad \mbox{and }
\tau_2^\delta = \grad \Psi_\delta\cdot \tau_2.$$ Using Proposition
\ref{diffeo}, we find that
\begin{equation}\label{vectortangent}
\tau_1^\delta = \tau_1+\delta M \tau_1,\quad \mbox{and }
\tau_2^\delta = \tau_2+\delta M \tau_2,
\end{equation}
where the $(3\times 3)$- matrix $M$ is given by
$$
M=\left(%
\begin{array}{ccc}
  \partial_1(\rho\nu_1) & \partial_2(\rho\nu_1) & \partial_3(\rho\nu_1) \\
  \partial_1(\rho\nu_2) & \partial_2(\rho\nu_2) & \partial_3(\rho\nu_2) \\
 \partial_1(\rho\nu_3) & \partial_2(\rho\nu_3) & \partial_3(\rho\nu_3) \\
\end{array}%
\right)
$$
with $\nu_i$ means the $i$-th ($i=1,2,3$) component of the vector $\nu$.\\

For $\delta$ sufficiently small, one can see that the outward unit
normal vector to $\partial \Omega_\delta$ is given by
\begin{equation}\label{vectornormal1} \nu_\delta(x) =
\frac{\tau_1^\delta(x) \times \tau_2^\delta(x)}{| \tau_1^\delta(x)
\times \tau_2^\delta(x)|},\end{equation} for $x\in \partial \Omega$.
Then, the following asymptotic expansion holds.

\begin{proposition}\label{lemnormal}Let $\nu_0$ be given by (\ref{normal0}). Then,  the outward unit normal $\nu_\delta(x)$ to $\partial\Omega_{\delta}$ at $x$, can be expanded uniformly as
\begin{equation} \label{nuexpan}\nu_\delta(x)=\nu^{(0)}(x)+\sum_{n=1}^{\infty} \delta^n
 \nu^{(n)}(x), \quad x \in \partial \Omega,
 \end{equation}
where the vector-valued functions $\nu^{(n)}$ are uniformly bounded
regardless of $n$. In particular, for $x\in \partial \Omega$
\begin{equation}\label{n0}
 \nu^{(0)}(x)=\nu(x),
\end{equation}
\begin{equation}\label{n1}
 \nu^{(1)}= \frac{1}{| \tau_1\times \tau_2|}\big[\tau_1\times M \tau_2+ M \tau_1\times\tau_2-\big(\nu_0\cdot( \tau_1\times M \tau_2+ M \tau_1\times\tau_2)  \big)\nu_0
\big].
\end{equation}
\end{proposition}
\proof

Considering the expansions (\ref{vectortangent}) for $\delta$
sufficiently small,  the  relation  (\ref{vectornormal1}) becomes
\begin{equation}\label{vectornormal2} \nu_\delta =  \frac{{\bf a}+\delta {\bf b}+\delta^2{\bf c}}{|{\bf a}+\delta {\bf b}+\delta^2{\bf c}|},\end{equation}
where  ${\bf a},  {\bf b}$, and ${\bf c}$  are vector-valued
functions  given  by

\begin{align*}
{\bf a}=\tau_1
\times \tau_2,\\
{\bf b}=  M \tau_1\times \tau_2 + \tau_1\times M \tau_2,\\
{\bf c}= M \tau_1\times M \tau_2.
\end{align*}

So that, by expanding the quotient~(\ref{vectornormal2})   as
$\delta$ tends to zero, we get the desired results.
\square\\

Next, one can use Proposition \ref{diffeo} to get a uniformly
convergent expansion for the surface element as follows:
\begin{proposition}\label{prop-surface} Let $\tilde{y}=\Psi_\delta (y)$ where $\Psi_\delta (y)$ is given by (\ref{psi-d}) for $y\in\partial\Omega$. Then,  the following expansion for the surface element $d\sigma_\delta(\tilde{y})$ holds uniformly for $y\in\partial\Omega$:
\begin{equation}\label{sigma-delta}d\sigma_\delta(\tilde{y})=\det \left( \nabla \Psi_\delta \right) d\sigma(y)=\big(\sigma_0(y)+\sigma_1(y)\delta+\sigma_2(y)\delta^2
+\sigma_3(y)\delta^3\big)d\sigma(y),\end{equation}
where $\sigma_0\equiv1$, $\sigma_1(y)=\nabla\cdot(\rho\nu)$, $\sigma_2(y)= \frac{1}{2}\big[\big(\rm{tr}  \nabla(\rho\nu)\big)^2-\rm{tr} \big(   \nabla (\rho \nu)^2  \big) \big] $, and $\sigma_3(y)=\det\left(  \nabla(\rho \nu)  \right)$.\\
\end{proposition}

Set
\begin{align}\label{eqx}
\tilde{x}=\Psi_{\delta}(x)&,\quad x\in\partial\Omega \\
\tilde{y}=\Psi_{\delta}(y)&,\quad y\in\partial\Omega\label{eqy},
\end{align}
and define
\begin{equation}\label{theta}
\Theta(x,y):=\frac{1}{\delta}\big(\Psi_{\delta}(x)-\Psi_{\delta}(y)-(x-
y)\big) \quad \mbox{for }\delta\neq 0.
\end{equation} \label{theta2}
Recall that   $\Psi_\delta$ is a $C^2$ vector-valued function on
$\partial\Omega$, then $\Theta$ is also a $C^2$  function on
$(\partial\Omega)^2$. Moreover, the following holds.
\begin{proposition}\label{Theta1}
The vector-valued function $\Theta(x,y)$  is $C^2$  on
$(\partial\Omega)^2$ and there exists a constant $C>0$ that only
depends on $\Omega$ and $\rho$  such that
\begin{align}\label{theta_reg}
|\Theta(x,y)|  \leq C |x-y|,\\
|\Theta(x,y)\cdot(x-y)|\leq C |x-y|^2,
\end{align}
for all $x, y \in \partial \Omega$.
 \end{proposition}
\proof Expression~(\ref{theta}) shows that
\begin{equation}
\Theta(x,y) \;=\; \rho(x)\nu(x) -\rho(y)\nu(y).
\end{equation}
Since $\partial \Omega$ is  a  $C^3$  surface,  there exists a
constant $C^\prime>0$ such that
\begin{align*}
|\nu(x) -\nu(y)|  \leq C^\prime |x-y|,\\
|\nu(x)\cdot (x-y)|\leq C^\prime|x-y|^2,
\end{align*}
for all $x, y \in \partial \Omega$. \\
 The last inequalities and the  $C^2$ regularity of   $\rho$ yield the results of the proposition.\\

\square\\

Now, by using (\ref{theta}) we get

\begin{equation}\label{eq1*}\tilde{x}-\tilde{y}=x-y+\delta\Theta(x,y),\quad (x,y)\in\partial\Omega\times \partial\Omega,
\end{equation}
and the following results hold.
\begin{lemma}\label{lemexpansion1} Let $r>0$ be a fixed real and $m \geq 2$ be a fixed integer.
The  following asymptotic expansions
\begin{eqnarray}\label{eq3}
\frac{e^{i\lambda|\tilde{x}-\tilde{y}|}}{|\tilde{x}-\tilde{y}|^m}\,=\,
\frac{e^{i\lambda|{x}-{y}|}}{|{x}-{y}|^m} \Big(1 +\delta
T_1^{(m)}(x,y)+\sum_{n\geq2}\delta^nT_n^{(m)}(\lambda;x,y)\Big),\\
\label{eq3*}
\partial_{\tilde x_i }\partial_{\tilde{x}_j}\frac{e^{i\lambda|\tilde{x}-\tilde{y}|}-1}
{|\tilde{x}-\tilde{y}|^3}\,=\, R_{0}^{(ij)}(\lambda;x,y) +\delta
R_{1}^{(ij)}(\lambda;x,y)+\sum_{n\geq2}\delta^nR_{n}^{(ij)}(\lambda;x,y)
\end{eqnarray}
hold uniformly for $(\lambda,x,y)\in B_r(0)\times\partial\Omega^2$,
where $B_r(0)$ is a ball in the complex plane  of center zero and
radius $r$. In addition the functions $T_n^{(m)}(\lambda;x,y)$ and $
R_{n}^{(ij)}(\lambda;x,y)$ are smooth and bounded uniformly on
$B_r(0)\times\partial\Omega^2$. The first coefficients are given by:
$$
R_{0}^{(ij)}(\lambda;x,y):=\partial_{x_i}\partial_{x_j}\frac{e^{i\lambda|x-y|}-1}
{|x-y|^3},$$

and

$$R_{1}^{(ij)}(\lambda;x,y)=(x-y)\cdot\Theta(x,y)\Big(-\lambda^2\delta_{ij}-3i\lambda \frac{\delta_{ij}}{r}-i\lambda^3\frac{r_ir_j}{r}
+7\lambda^2
\frac{r_ir_j}{r^2}+15i\lambda\frac{r_ir_j}{r^3}\Big)\frac{e^{i\lambda
r}}{r^5}$$

$$
+(x-y)\cdot\Theta(x,y)\Big(4i\lambda\frac{\delta_{ij}}{r}-15\frac{\delta_{ij}}{r^2}-\lambda^2[-5\frac{r_ir_j}{r^2}+
\frac{(r_i\theta_j+r_j\theta_i)}{(x-y)\cdot\Theta(x,y)} -7i\lambda
[-6\frac{r_ir_j}{r^3}+\frac{1}{r}\frac{(r_i\theta_j+r_j\theta_i)}{(x-y)\cdot\Theta(x,y)}]$$
$$
+15[-7\frac{r_ir_j}{r^4}+\frac{1}{r^2}\frac{(r_i\theta_j+r_j\theta_i)}{(x-y)\cdot\Theta(x,y)}\Big)\frac{e^{i\lambda
 r}}{r^5}$$
 $$
 +15\frac{(x-y)\cdot\Theta(x,y)}{r^7}\Big(\delta_{ij}-[-7\frac{r_ir_j}{r^2}+\frac{(r_i\theta_j+r_j\theta_i)}{(x-y)\cdot\Theta(x,y)}]\Big),
$$
where $r=|x-y|$, $r_j=x_j-y_j$, $\theta_j$ means the $j^{\mbox{th}}$
component of $\Theta$, and $\delta_{ij}$ means the Kronecker index.
\end{lemma}
\proof For $m= 2$  and $x\not=y$ we have
$$|\tilde{x}-\tilde{y}|^2=|x-y+\delta\Theta(x,y)|^2
\;=\;|x-y|^2\left(1+ \delta c_1^{(2)}(x,y) + \delta^2 c_2^{(2)}(x,y)
\right),$$ where
\begin{eqnarray*}
  c_1^{(2)}(x,y) \;=\; \frac{2\Theta(x,y)\cdot (x-y)}{|x-y|^2}, &
c_2^{(2)}(x,y) \;=\; \frac{\Theta(x,y)\cdot\Theta(x,y)}{|x-y|^2}.
\end{eqnarray*}
Proposition~\ref{theta2} shows that $ c_1^{(2)}(x,y)$ and $
c_2^{(2)}(x,y)$ are bounded uniformly on $ \partial \Omega^2$. \\

For $m>2$, we have:
$$
|\tilde{x}-\tilde{y}|^m=|x-y+\delta\Theta(x,y)|^m=|x-y|^m\big|1+\delta\frac{\Theta(x,y)}{|x-y|}\big|^m,$$where
$x\neq y$.\\
Using (\ref{theta}), Proposition \ref{diffeo} and Proposition
\ref{Theta1}, we see that the regular vector-valued function
$(x,y)\mapsto\frac{\Theta(x,y)}{|x-y|}$
is well defined on $\partial\Omega\times\partial\Omega,$ and it is independent of $\delta$.\\
Therefore, we can expand
\begin{equation}\label{eq-x1}|\tilde{x}-\tilde{y}|^m=c_0^{(m)}(x,y)+\sum_{n=1}^{\infty}\delta^n c_n^{(m)}(x,y)\quad
 \mbox{uniformly on } \partial\Omega\times \partial\Omega,
\end{equation}
where the first coefficients $c_0^{(m)}(x,y)=|x-y|^m$,
$c_1^{(m)}(x,y)=m<x-y,\Theta(x,y)>|x-y|^{m-2}$ and
$c_2^{(m)}(x,y)=\frac{m}{2}|x-y|^{m}\big[|\frac{\Theta(x,y)}{|x-y|}|^2+(m-2)\big(<\frac{x-y}{|x-y|^2},\frac{\Theta(x,y)}{|x-y|}>\big)^2
\big].$ Moreover, if $m$ is even, then $c_n^{(m)}(x,y)=0$ for $n\geq
m+1$.\\ Now combining (\ref{eq-x1}) for $m=1$ with the well known
asymptotic expansion of the exponential function, we immediately get
\begin{equation}\label{eq-x2}
e^{i\lambda|\tilde{x}-\tilde{y}|}=\ds
\sum_{n=0}^{\infty}\delta^nK_n(\lambda;x,y)\quad \mbox{uniformly on
}B_r(0)\times\partial\Omega\times\partial\Omega,
\end{equation}
where the first coefficients $K_0(\lambda;x,y)=e^{i\lambda
c_{0}^{(1)}}=e^{i\lambda |x-y|}$, and the coefficients
$K_n(\lambda;x,y)$ can be deduced recursively from $c_n^{(1)}$.\\
Thanks to relations (\ref{eq1*}), (\ref{eq-x1}) and (\ref{eq-x2}),
we can obtain the desired result given by (\ref{eq3}) where the
coefficients $T_n^{(m)}$  can be deduced easily from $c_n^{(m)}$ and
$K_n$.\\

To prove relation (\ref{eq3*}), we firstly expand
\begin{equation}\label{eq-expand1}
\partial_{\tilde x_i }\partial_{\tilde{x}_j}\frac{e^{i\lambda|\tilde{x}-\tilde{y}|}-1}
{|\tilde{x}-\tilde{y}|^3}=\frac{1}{|\tilde{x}-\tilde{y}|^3}\partial_{\tilde
x_i
}\partial_{\tilde{x}_j}\big(e^{i\lambda|\tilde{x}-\tilde{y}|}-1\big)+\partial_{\tilde
x_i
}\big(\frac{1}{|\tilde{x}-\tilde{y}|^3}\big)\partial_{\tilde{x}_j}\big(e^{i\lambda|\tilde{x}-\tilde{y}|}-1\big)
\end{equation}
$$+\partial_{\tilde
x_j
}\big(\frac{1}{|\tilde{x}-\tilde{y}|^3}\big)\partial_{\tilde{x}_i}\big(e^{i\lambda|\tilde{x}-\tilde{y}|}-1\big)+\partial_{\tilde
x_i
}\partial_{\tilde{x}_j}\big(\frac{1}{|\tilde{x}-\tilde{y}|^3}\big)\big(e^{i\lambda|\tilde{x}-\tilde{y}|}-1\big).
$$
To simplify, let us denote $r=|x-y|$, $r_i=x_i-y_i$,
$\tilde{r}=|\tilde{x}-\tilde{y}|$, and $\hat{r}_{j}$ defined by the
relation
\begin{equation}\label{eq-expand2}\partial_{\tilde{x}_j}\tilde{r}=\frac{\hat{r}_{j}}{\tilde{r}}.
\end{equation}
Then,
\begin{equation}\label{eq-expand3}\partial_{\tilde{x}_j}(\frac{1}{\tilde{r}^3})=-3\frac{\hat{r}_{j}}{\tilde{r}^5},\quad
\mbox{and
}\partial_{\tilde{x}_i}\partial_{\tilde{x}_j}(\frac{1}{\tilde{r}^3})=-3\frac{\delta_{ij}}{\tilde{r}^5}+15\frac{\hat{r}_i\hat{r}_j}{\tilde{r}^7}.
\end{equation}

Now, by using the following result
\begin{equation}\label{eq-expand4}
\partial_{\tilde{x}_j}\big(e^{i\lambda|\tilde{x}-\tilde{y}|}-1\big)=\partial_{\tilde{x}_j}(i\lambda\tilde{r}\big)e^{i\lambda\tilde{r}}
=i\lambda\frac{\hat{r}_j}{\tilde{r}}e^{i\lambda\tilde{r}},
\end{equation}
we get that\begin{equation}\label{eq-expand5}
\partial_{\tilde{x}_i}\partial_{\tilde{x}_j}\big(e^{i\lambda|\tilde{x}-\tilde{y}|}-1\big)=
i\lambda\big[
\partial_{\tilde{x}_i}(\frac{\hat{r}_j}{\tilde{r}})e^{i\lambda\tilde{r}}+\frac{\hat{r}_j}{\tilde{r}}\partial_{\tilde{x}_i}e^{i\lambda\tilde{r}}
\big]=i\lambda\big[\frac{\delta_{ij}}{\tilde{r}} -
\frac{\hat{r}_i\hat{r}_j}{\tilde{r}^3} +i\lambda
\frac{\hat{r}_i\hat{r}_j}{\tilde{r}^2} \big]e^{i\lambda\tilde{r}}.
\end{equation}
To find the desired result in (\ref{eq3*}), we may use
(\ref{eq-expand3}), (\ref{eq-expand4}) and (\ref{eq-expand5}) to see
that the relation (\ref{eq-expand1}) verifies:
\begin{equation}\label{eq-expand6}
\partial_{\tilde x_i }\partial_{\tilde{x}_j}\frac{e^{i\lambda|\tilde{x}-\tilde{y}|}-1}
{|\tilde{x}-\tilde{y}|^3}=\big[
i\lambda\frac{\delta_{ij}}{\tilde{r}^4}-3\frac{\delta_{ij}}{\tilde{r}^5}-\lambda^2\frac{\hat{r}_i\hat{r}_j}{\tilde{r}^5}
-7i\lambda\frac{\hat{r}_i\hat{r}_j}{\tilde{r}^6}+15\frac{\hat{r}_i\hat{r}_j}{\tilde{r}^7}
\big]e^{i\lambda\tilde{r}}+3\frac{\delta_{ij}}{\tilde{r}^5}-15\frac{\hat{r}_i\hat{r}_j}{\tilde{r}^7}.
\end{equation}

On the other hand, the components of the vectorial relation
(\ref{eq1*}) can be given as follows
\begin{equation}\label{eq4}\tilde{x}_i-\tilde{y}_i=x_i-y_i+\delta\theta_i(x,y).\quad i=1,2,3 \end{equation}
where $\theta_i(x,y)$ means the $i^{\mbox{th}}$ component of the
vector-valued function $\Theta(x,y)$.\\ Then, by relations
(\ref{eq-expand2}) and (\ref{eq4}) we deduce that
\begin{equation}\label{eq-expand7}\hat{r}_j=\tilde{x}_i-\tilde{y}_i=x_i-y_i+\delta\theta_i(x,y)=
r_i+\delta\theta_i(x,y),\quad i=1,2,3.\end{equation} Using both
relations (\ref{eq1*}) and (\ref{eq-expand7}), we get the following
expansion
\begin{equation}\label{eq-expand8}\hat{r}_i\hat{r}_j=\alpha_{0}^{(ij)}+\alpha_{1}^{(ij)}\delta+\alpha_{2}^{(ij)}\delta^2
,\end{equation} where the first coefficients:
$\alpha_{0}^{(ij)}=r_ir_j,$
$\alpha_{1}^{(ij)}=r_i\theta_j+r_j\theta_i$ and
$\alpha_{2}^{(ij)}=\theta_i\cdot\theta_j.$\\

Now regarding (\ref{eq-x1}) and using the fact that
$c_{0}^{(m)}\neq0$ for each integer $m$. Then one can expand
\begin{equation}\label{eq-expand9}\displaystyle \frac{1}{\tilde{r}^m}=\kappa_0^{(m)}(x,y)+\sum_{n=1}^\infty\delta^n\kappa_{n}^{(m)}(x,y)
\quad \mbox{uniformly },
\end{equation}
where the first coefficients:
$\kappa_0^{(m)}(x,y)=(c_{0}^{(m)})^{-1}(x,y)$, and
$\kappa_1^{(m)}(x,y)=-\big(c_{1}^{(m)}(c_{0}^{(m)})^{-2}\big)(x,y)$.
By using (\ref{eq-expand8})-(\ref{eq-expand9}), we get that
\begin{equation}\label{eq-expand10}\displaystyle \frac{\hat{r}_i\hat{r}_j}{\tilde{r}^m}=
\beta_{0,m}^{(ij)}+\delta\beta_{1,m}^{(ij)}+\sum_{n=2}^\infty\delta^n\beta_{n,m}^{(ij)},
\end{equation}
where the first coefficients:
$\beta_{0,m}^{(ij)}(x,y)=\alpha_0^{(ij)}(c_{0}^{(m)})^{-1}(x,y)$,
and\\
$\beta_{1,m}^{(ij)}(x,y)=-\alpha_0^{(ij)}\big(c_{1}^{(m)}(c_{0}^{(m)})^{-2}\big)(x,y)+\alpha_1^{(ij)}(c_{0}^{(m)})^{-1}(x,y)$.\\
To achieve the proof, we may insert all expansions (\ref{eq-x2}),
(\ref{eq-expand9}) (for $m\in\{4,5\}$), and (\ref{eq-expand10})(for
$m\in\{5,6,7\}$ into (\ref{eq-expand6}). We get that
$$
\partial_{\tilde x_i }\partial_{\tilde{x}_j}\frac{e^{i\lambda|\tilde{x}-\tilde{y}|}-1}
{|\tilde{x}-\tilde{y}|^3}=R_{0}^{(ij)}(x,y)+\delta
R_{1}^{(ij)}(x,y)+\sum_{n=2}^{\infty}\delta^nR_{n}^{(ij)}(x,y)$$
where the first coefficient:
$$R_{0}^{(ij)}(\lambda;x,y)=\big(i\lambda\delta_{ij}\kappa_0^{(4)}-3\delta_{ij}\kappa_0^{(5)}-\lambda^2\beta_{0,5}^{(ij)}-7i\lambda\beta_{0,6}^{(ij)}
+15\beta_{0,7}^{(ij)}\big)K_0+3\delta_{ij}\kappa_{0}^{(5)}-15\beta_{0,7}^{(ij)}$$
$$=\Big[i\lambda\delta_{ij}(c_{0}^{(4)})^{-1}-3\delta_{ij}(c_{0}^{(5)})^{-1}-\lambda^2r_ir_j(c_{0}^{(5)})^{-1}
-7i\lambda r_ir_j(c_{0}^{(6)})^{-1}
+15r_ir_j(c_{0}^{(7)})^{-1}\Big]e^{i\lambda c_{0}^{(1)}}$$
$$+3\delta_{ij}(c_{0}^{(5)})^{-1}-15r_ir_j(c_{0}^{(7)})^{-1}.$$
Using the fact that $c_{0}^{(m)}=|x-y|^m$, we find that $$
R_{0}^{(ij)}(\lambda;x,y):=\partial_{x_i}\partial_{x_j}\frac{e^{i\lambda|x-y|}-1}
{|x-y|^3}.$$

Based on (\ref{eq-expand6}), we find that
$$R_{1}^{(ij)}(\lambda;x,y)=\big(i\lambda\delta_{ij}\kappa_0^{(4)}-3\delta_{ij}\kappa_0^{(5)}-\lambda^2\beta_{0,5}^{(ij)}-7i\lambda\beta_{0,6}^{(ij)}
+15\beta_{0,7}^{(ij)}\big)K_1$$
$$
+\big(i\lambda\delta_{ij}\kappa_1^{(4)}-3\delta_{ij}\kappa_1^{(5)}-\lambda^2\beta_{1,5}^{(ij)}-7i\lambda\beta_{1,6}^{(ij)}
+15\beta_{1,7}^{(ij)}\big)K_0+3\delta_{ij}\kappa_{1}^{(5)}-15\beta_{1,7}^{(ij)}.
$$
So that, by using the fact that
$K_1=i\lambda\Theta\cdot\frac{(x-y)}{r}e^{i\lambda c_{0}^{(1)}}$, we
get
$$R_{1}^{(ij)}(\lambda;x,y)=i\lambda\Theta\cdot\frac{(x-y)}{r}\Big(i\lambda\delta_{ij}(c_{0}^{(4)})^{-1}-3\delta_{ij}(c_{0}^{(5)})^{-1}-\lambda^2r_ir_j(c_{0}^{(5)})^{-1}
-7i\lambda r_ir_j(c_{0}^{(6)})^{-1}$$
$$
+15r_ir_j(c_{0}^{(7)})^{-1}\Big)e^{i\lambda c_{0}^{(1)}}$$
$$
+\Big(i\lambda\delta_{ij}(c_{0}^{(4)})^{-2}c_{1}^{(4)}-3\delta_{ij}(c_{0}^{(5)})^{-2}c_{1}^{(5)}-\lambda^2[-r_ir_j(c_{0}^{(5)})^{-2}
c_{1}^{(5)}+(r_i\theta_j+r_j\theta_i)(c_{0}^{(5)})^{-1} ]$$
$$-7i\lambda [-r_ir_j(c_{0}^{(6)})^{-2}
c_{1}^{(6)}+(r_i\theta_j+r_j\theta_i)(c_{0}^{(6)})^{-1} ]
+15[-r_ir_j(c_{0}^{(7)})^{-2}
c_{1}^{(7)}+(r_i\theta_j+r_j\theta_i)(c_{0}^{(7)})^{-1}
]\Big)e^{i\lambda c_{0}^{(1)}}$$
$$
+3\delta_{ij}(c_{0}^{(5)})^{-2}c_{1}^{(5)}-15[-r_ir_j(c_{0}^{(7)})^{-2}
c_{1}^{(7)}+(r_i\theta_j+r_j\theta_i)(c_{0}^{(7)})^{-1} ].
$$
Using the explicit forms of the coefficients $c_{0}^{(m)}$ and
$c_{1}^{(m)}$ given above, we immediately get the desired expression
of $R_{1}^{(ij)}(\lambda;x,y)$. The other coefficients can be
deduced easily by the same manner in terms of $c_{n}^{(m)}$.
\square\\

Now using (\ref{eq4}), we obtain for $i,j=1,2,3$ that:
\begin{equation}\label{eq5}(\tilde{x}_i-\tilde{y}_i)(\tilde{x}_j-\tilde{y}_j)=\hat{g}_0(x,y)+\delta\hat{g}_1(x,y)+\delta^2\hat{g}_2(x,y),
\end{equation}
where
$$
\hat{g}_0(x,y)=(x_i-y_i)(x_j-y_j),\quad
\hat{g}_1(x,y)=\theta_i(x,y)(x_j-y_j)+\theta_j(x,y)(x_i-y_i),$$ and
$$\hat{g}_2(x,y)=\theta_i(x,y)\cdot \theta_j(x,y).$$

Now, by using  (\ref{sigma-delta}), (\ref{eq-x1}) and (\ref{eq5}),
we immediately get
\begin{equation}\label{eq6}
\ds\frac{(\tilde{x}_i-\tilde{y}_i)(\tilde{x}_j-\tilde{y}_j)}{|\tilde{x}-
\tilde{y}|^{m}}d\sigma_\delta(\tilde{y})=\Big(\hat{T}_0(x,y)+\delta
\hat{T}_1(x,y)+\sum_{n\geq2}\delta^n\hat{T}_n(x,y)\Big)d\sigma(y),
\end{equation}
where the Taylor coefficients $\hat{T}_n$ can be given explicitly with the aid of $\hat{g}_0,\hat{g}_1$, and $\hat{g}_2$.\\

Next, the following result holds.

\begin{lemma}\label{lemexpansion2}
The following uniform expansion holds on
$\R\times\partial\Omega\times\partial\Omega$:
\begin{equation}\label{eq7}
\ds
(\tilde{x}_i-\tilde{y}_i)(\tilde{x}_j-\tilde{y}_j)|\tilde{x}-\tilde{y}|e^{i\lambda|\tilde{x}-\tilde{y}|}d\sigma_\delta(\tilde{y})=
E_\delta(\lambda;x,y)~d\sigma(y)=\sum_{n=0}^{\infty}\delta^nE_n(\lambda;x,y)~d\sigma(y)
\end{equation}
with
$$
E_0(\lambda;x,y)=\hat{g}_0(x,y)|x-y|e^{i\lambda|x-y|},
$$and the other coefficients $E_n$ are deduced from those $\sigma_n$, $c_n^{(1)}$, $K_n$ and $\hat{g}_n.$
\end{lemma}
\proof From  the proof of Lemma \ref{lemexpansion1} (for $m=1$) and
from relations (\ref{sigma-delta}) and (\ref{eq5}), one can get that
$$
\ds
(\tilde{x}_i-\tilde{y}_i)(\tilde{x}_j-\tilde{y}_j)|\tilde{x}-\tilde{y}|e^{i\lambda|\tilde{x}-\tilde{y}|}d\sigma_\delta(\tilde{y})=\Big(\hat{g}_0(x,y)+\delta\hat{g}_1(x,y)+
$$
$$
\delta^2\hat{g}_2(x,y)\Big)\Big(\sum_{n=0}^{\infty}\delta^n
c_{n}^{(1)}  \Big)\Big(\sum_{n=0}^{\infty}K_n(\lambda;x,y)
\Big)\Big(\sigma_0(y)+\delta\sigma_1(y)+$$
$$\delta^2\sigma_2(y)+\delta^3\sigma_3(y)\Big)~d\sigma(y).$$
By collecting terms of equal powers in the above relation, one can
deduce easily the uniform expansion (\ref{eq7}) with
$$
E_0(\lambda;x,y)=\hat{g}_0(x,y)|x-y|e^{i\lambda|x-y|}.
$$
\square\\

Now, from (\ref{system3}) we have
$$
\Gamma_{ij}(\lambda,|x-y|)=\frac{e^{i\lambda
r}}{4\pi}\big[\frac{\delta_{ij}}{r}-\frac{\hat{x}_i\hat{x}_j}{r^3}
\big]+ \frac{e^{i\lambda r}}{4\pi
\lambda}\big[i\frac{\delta_{ij}}{r^4}-4i\frac{\hat{x}_i\hat{x}_j}{r^4}-3ir\hat{x}_i\hat{x}_j
\big]$$
$$+
\frac{1}{
4\pi\lambda^2}\big[-3\frac{\delta_{ij}}{r^5}+15\frac{\hat{x}_i\hat{x}_j}{r^7}
\big] +\frac{e^{i\lambda r}}{4\pi
\lambda^2}\big[-3\frac{\delta_{ij}}{r^5}+15\frac{\hat{x}_i\hat{x}_j}{r^3}
\big],
$$
where $r=|x-y|$ and $\hat{x}_i=x_i-y_i$. Then, by inserting
relations (\ref{eq3}), (\ref{eq5}), (\ref{eq6}) and that given by
Lemma \ref{lemexpansion2} into above identity, we immediately get
the following main results

\begin{proposition}\label{asym-Gamma}Let the perturbed boundary $\partial\Omega_\delta$ defined by (\ref{parametr}). Let $\tilde x$ and $\tilde y$ given by (\ref{eqx})-(\ref{eqy}), and the surface element $d\sigma_\delta(\tilde{y})$ given by (\ref{sigma-delta}). Then, the components $\Gamma_{ij},1\leq i,j\leq 3$ of the fundamental Stokes tensor can be expanded uniformly as:
\begin{equation}\label{eq8}
\Gamma_{ij}(\lambda,|\tilde{x}-\tilde{y}|)~d\sigma_\delta(\tilde{y})=\Big(\Gamma_{ij}^{(0)}(\lambda,|x-y|)+\delta\Gamma_{ij}^{(1)}(\lambda,|x-y|)\end{equation}
$$+\sum_{n\geq2}\delta^n\Gamma_{ij}^{(n)}(\lambda,|x-y|)\Big)~d\sigma(y),\quad y\in\partial\Omega
$$
where the first coefficients:
$$
\Gamma_{ij}^{(0)}(\lambda,|x-y|):=\frac{\delta_{ij}}{4\pi}\Big(T_{0}^{(1)}+\frac{i}{\lambda}
T_{0}^{(4)}-3\frac{r_0}{\lambda^2}-3\frac{T_{0}^{(5)}}{\lambda^2}\Big)+
\frac{1}{4\pi}(\frac{15}{\lambda^2}-1)T_{0}^{(3)}\hat{g}_0-
\frac{4i}{4\pi\lambda}T_{0}^{(4)}\hat{g}_0$$
$$ - \frac{3i}{4\pi\lambda}E_0+\frac{15}{4\pi\lambda^2}\hat{T}_0,$$ and

$$\Gamma_{ij}^{(1)}(\lambda,|x-y|):=\frac{\delta_{ij}}{4\pi}\Big(T_{1}^{(1)}+\frac{i}{\lambda} T_{1}^{(4)}-3\frac{r_1}{\lambda^2}-3\frac{T_{1}^{(5)}}{\lambda^2}\Big)+
\frac{1}{4\pi}(\frac{15}{\lambda^2}-1)\big[T_{0}^{(3)}\hat{g}_1+T_{1}^{(3)}\hat{g}_0\big]$$
$$ - \frac{4i}{4\pi\lambda}\big[T_{0}^{(4)}\hat{g}_1+T_{1}^{(4)}\hat{g}_0 \big]-
\frac{3i}{4\pi\lambda}E_1+\frac{15}{4\pi\lambda^2}\hat{T}_1 .$$
\end{proposition}
In Proposition \ref{asym-Gamma}, the coefficients $r_0$ and $r_1$ are deduced from (\ref{eq-x1}) for $m=5$.\\

Now, introduce a sequence of components of integral operators
$(\mathcal{A}_{i}^{(n)})_{n\geq0}$,  defined for any $\varphi\in
L^2(\partial\Omega)^3$ by:
\begin{equation}\label{Sn}
\ds
\big(\mathcal{A}^{(n)}\varphi\big)_{i}(x)=\int_{\partial\Omega}\Gamma_{ij}^{(n)}(x,y)\varphi_j(y)~d\sigma(y),\quad
\mbox{for }i,j\in\{1,2,3\} \mbox{ and } n\geq 0.
\end{equation}
Using previous results, il is clear that we know explecitely the
first terms $\mathcal{A}_{i}^{(0)}$, $\mathcal{A}_{i}^{(1)}$. For
any positive integer $N$, we can by recursive method get the term
$\mathcal{A}_{i}^{(N)}$. Then, the following theorem holds.
\begin{theorem}\label{estimation-op1} Let $\mathcal{A}^{\delta}$ be the operator defined by (\ref{op1}). Let $\Psi_{\delta}(x)$ be the diffeomorphism defined as in Lemma \ref{diffeo}. Let $N$ be a positive integer. There exists a positive constant $C$ depending only on
$N$, and $\| \rho \|_{C^2}$ such that for any $\tilde\varphi \in
L^2(\partial {\Omega}_{\delta})^3$ and $i,j\in\{1,2,3\}$, the
$i^{th}$-component $\mathcal{A}_{i}^{\delta}$ defined by
(\ref{op1*}) satisfies:
$$
\big\|\big(\mathcal{A}^{\delta}\tilde{\varphi}\big)_{i}
o\Psi_{\delta}-\big(\mathcal{A}^{(0)}\varphi)_{i}-\sum_{n=1}^{N}\delta^n\big(\mathcal{A}^{(n)}
\varphi\big)_{i}\big\|_{L^2(\partial \Omega)}\leq
C\delta^{N+1}\|\varphi\|_{L^2(\partial \Omega)^3},
$$
where $\varphi=\tilde{\varphi}o\Psi_{\delta}$.
\end{theorem}

\subsection{Asymptotic expansion of the eigenelements}
To develop asymptotic behaviors of eigenvalues and eigenfunctions
with respect to the parameter of perturbation $\delta$, we may use
the results of Theorem \ref{estimation-op1}. Then,  the following
asymptotic expansion related to the operator $\mathcal{A}_\delta$
appears clearly.
\begin{proposition} Suppose that we have all hypothesis of Theorem \ref{estimation-op1}. Then,  the operator $\mathcal{A}^\delta(\lambda)$ defined by (\ref{op1})
can be expanded uniformly for $x\in\partial\Omega$ as follows:
\begin{equation}\label{expan1}
\mathcal{A}^\delta(\lambda)\varphi=\mathcal{A}^{(0)}(\lambda)\varphi+\delta\mathcal{A}^{(1)}(\lambda)\varphi+\delta^2\mathcal{A}^{(2)}(\lambda)\varphi
+\cdots;\quad \mbox{as }\delta\to0,
\end{equation}
where $\varphi\in L^2(\partial\Omega)^3$, the $i^{th}$-component of
the first term is given by
$$
\big(\mathcal{A}_{i}^{(0)}(\lambda)\varphi\big)_i(x)=\int_{\partial\Omega}\Gamma_{ij}^{(0)}(x,y)\varphi_j(y)~d\sigma(y),\quad
\mbox{for }i,j\in\{1,2,3\},
$$
and more generally, the $i^{th}$-component of the term with order
$n$ is given by
$$\big(
\mathcal{A}^{(n)}(\lambda)\varphi\big)_i(x)=\int_{\partial\Omega}\Gamma_{ij}^{(n)}(x,y)\varphi_j(y)~d\sigma(y),\quad
\mbox{for }i,j\in\{1,2,3\} \mbox{ and } n\geq 1.
$$
The coefficients $\Gamma_{ij}^{(n)}$ are given by (\ref{eq8}).
\end{proposition}

Let  $a_{j}(\delta)$ denotes: \begin{equation}\label{a-delta}
a_{j}(\delta) = \frac{1}{2i\pi}tr\int_{\partial{D_{\epsilon_{0}}}}
(\lambda-\lambda_{0})^j (\mathcal{A}^{\delta})^{-1}( \lambda)
\partial_{\lambda} \mathcal{A}^{\delta}( \lambda)\varphi
d\lambda,\end{equation}where $\varphi\in L^2(\partial\Omega)^3$. The
functions $a_{j}(\delta)$ is analytic in a complex neighborhood of
$0$ and satisfies: $\overline{a_{j}(\delta)} =
a_{j}(\overline{\delta}) $. The following main result holds.

\begin{proposition} \label{th4} Suppose that $\lambda_0$ is an
eigenvalue with multiplicity 1 of the eigenvalue problem
(\ref{system1}). Let the operator $\mathcal{A}^{\delta}$ be defined
by (\ref{op1}). Then, there exists a small positive number
$\delta_0$ such that the eigenvalue $\lambda(\delta)$ is analytic in
$ ]-\delta_{0},\delta_{0}[$ and satisfy:
 \begin{equation}\label{asympt-vp}\lambda(\delta)=\ds \lambda_{0} +
 \delta\lambda_{1}+\sum_{n\geq2}\lambda_{n}\delta^{n},\end{equation}
where the first coefficients are given explicitly by:
$$ \left\{
\begin{array}{lll}
\lambda_1=\ds \frac{1}{2i\pi}\mbox{ tr
}\int_{\partial{D_{\epsilon_{0}}}}(\lambda-\lambda_0)\big[
(\mathcal{A}^{(0)})^{-1}( \lambda)\mathcal{A}^{(1)}(
\lambda)(\mathcal{A}^{(0)})^{-1}(
\lambda)\partial_\lambda\mathcal{A}^{(0)}( \lambda) \big]\varphi~d\lambda,\\
\nm \lambda_2=\ds \frac{1}{2i\pi}\mbox{ tr
}\int_{\partial{D_{\epsilon_{0}}}}(\lambda-\lambda_0)\big[
(\mathcal{A}^{(0)})^{-1}\mathcal{A}^{(1)}(\mathcal{A}^{(0)})^{-1}\partial_\lambda\mathcal{A}^{(1)}
+
(\mathcal{A}^{(0)})^{-1}\mathcal{A}^{(2)}(\mathcal{A}^{(0)})^{-1}\partial_\lambda\mathcal{A}^{(0)}
+\\
\nm \ds
(\mathcal{A}^{(0)})^{-1}\big(\mathcal{A}^{(1)}(\mathcal{A}^{(0)})^{-1}\big)^2\partial_\lambda\mathcal{A}^{(0)}
\big]\varphi~d\lambda,
\end{array} \right.
$$
where $\varphi\in L^2(\partial\Omega)^3$
\end{proposition}
\proof Let $\lambda_\delta$ be the eigenvalue of the eigenvalue
problem (\ref{system2}) and let $\varphi\in
L^2(\partial\Omega)^3$.\\
Then, if we take the curl of the first equation in (\ref{system2}),
we see that there exists a function $w=w(v_\delta)$ called vorticity
associated to $v_\delta$ such that
\begin{equation}\label{vorticity1}
 \begin{cases}
   \Delta w+\lambda_\delta w=0 & \mbox{in }\Omega\\
  v=0 & \mbox{in }\partial\Omega. \\
  \end{cases}
\end{equation}
That is, $w$ is an eigenfunction of the negative Laplacian, but with
boundary conditions on the velocity $v_\delta$.\\
Next, let $u_\delta$ be the stream function for $v_\delta$ given as
in  Lemma 2.10 of \cite{james}. Then $w=\Delta u_\delta$ and $\nabla
u_\delta=0$ on $\partial\Omega$. Since $u_\delta$ is determined only
up to a constant we can then assume that $u_\delta=0$ on
$\partial\Omega$.\\ Thus, $u_\delta$ satisfies the following
eigenvalue problem for the Dirichlet biharmonic operator:
\begin{equation}\label{vorticity2}
 \begin{cases}
  \Delta \Delta u_\delta+\lambda_\delta \Delta u_\delta=0 & \mbox{in }\Omega\\
  u_\delta=0 & \mbox{in }\partial\Omega. \\
  \end{cases}
\end{equation}
Note that Temam \cite{temam2} exploits the similar correspondence
between the Stokes problem and the biharmonic problem in the proof
of the regularity of solutions to the Stokes system and to justify
several results. Moreover, as pointed out by Ashbaugh in
\cite{ashbaugh}, there is a similar correspondence between the
eigenvalue problems for the Dirichlet Laplacian and system
(\ref{vorticity2}) with the boundary condition $\nabla \cdot
u_\delta=0$ replaced by $\Delta u_\delta=0$.\\ Then, one can exploit
this correspondence to use the approach used in \cite{AT} to develop
an asymptotic expansion for the
eigenvalue.\\

So that, it is well known that there exits a polynomial-valued
function $\delta\mapsto \mathcal{Q}_\delta(\lambda)$ of degree 1,
analytic in $]-\delta_0,\delta_0[$ and of the
form:$$\mathcal{Q}_\delta(\lambda)=\lambda-a_1(\delta)$$such that
the perturbation $\lambda_\delta-\lambda_0$ is precisely its zero.
For the existence of $\mathcal{Q}_\delta$ one can follow the general
approach used, for example, in \cite{AT} for the case of Laplace
operator.\\ Writing:
$$\mathcal{Q}_\delta(\lambda_\delta-\lambda_0)=0.$$ Then we have
$$
\lambda_\delta-\lambda_0=a_1(\delta).
$$
Therefore, by (\ref{a-delta}) we have
\begin{equation}\label{a-delta-1}
\lambda_\delta-\lambda_0=
\frac{1}{2i\pi}tr\int_{\partial{D_{\epsilon_{0}}}}
(\lambda-\lambda_{0}) (\mathcal{A}^{\delta})^{-1}( \lambda)
\partial_{\lambda} \mathcal{A}^{\delta}( \lambda)\varphi
d\lambda.\end{equation} On the other hand, for $\delta$ in a small
neighborhood of $0$, the following Neumann series converges
uniformly with respect to $\lambda$ in $\partial D_{\epsilon_0}$:
$$(\mathcal{A}^{\delta})^{-1}( \lambda)=(\mathcal{A}^{(0)})^{-1}(
\lambda)+\sum_{k=1}^{\infty}(\mathcal{A}^{(0)})^{-1}( \lambda)\big[
 \big(\mathcal{A}^{(0)}( \lambda)- \mathcal{A}^{\delta}( \lambda)\big)(\mathcal{A}^{(0)})^{-1}(
\lambda)  \big]^k.
$$
So
that,$$\lambda_\delta-\lambda_0=\frac{1}{2i\pi}tr\int_{\partial{D_{\epsilon_{0}}}}
(\lambda-\lambda_{0}) (\mathcal{A}^{(0)})^{-1}(
\lambda)\partial_{\lambda} \mathcal{A}^{\delta}(
\lambda)\varphi~d\lambda +
$$
$$
\frac{1}{2i\pi}tr\int_{\partial{D_{\epsilon_{0}}}}
(\lambda-\lambda_{0}) \sum_{k=1}^{\infty}(\mathcal{A}^{(0)})^{-1}(
\lambda)\big[
 \big(\mathcal{A}^{(0)}( \lambda)- \mathcal{A}^{\delta}( \lambda)\big)(\mathcal{A}^{(0)})^{-1}(
\lambda)  \big]^k\partial_{\lambda} \mathcal{A}^{\delta}(
\lambda)\varphi~d\lambda
$$

By using (\ref{eq10}), we find that $\displaystyle
\frac{1}{2i\pi}tr\int_{\partial{D_{\epsilon_{0}}}}
(\lambda-\lambda_{0}) (\mathcal{A}^{(0)})^{-1}(
\lambda)\partial_{\lambda} \mathcal{A}^{\delta}(
\lambda)\varphi~d\lambda=0$. This result is a direct consequence of
the fact that $\mathcal{R}_0(\lambda)$ and $\partial_{\lambda}
\mathcal{A}^{\delta}( \lambda)$ are holomorphic in the variable
$\lambda$.\\

Now we have:$$\lambda_\delta-\lambda_0=
\frac{1}{2i\pi}tr\int_{\partial{D_{\epsilon_{0}}}}
(\lambda-\lambda_{0})(\mathcal{A}^{(0)})^{-1}( \lambda)\big[
 \big(\mathcal{A}^{(0)}( \lambda)- \mathcal{A}^{\delta}( \lambda)\big)(\mathcal{A}^{(0)})^{-1}(
\lambda)  \big]\partial_{\lambda} \mathcal{A}^{\delta}(
\lambda)\varphi~d\lambda+
$$
$$\frac{1}{2i\pi}tr\int_{\partial{D_{\epsilon_{0}}}}
(\lambda-\lambda_{0})(\mathcal{A}^{(0)})^{-1}( \lambda)\big[
 \big(\mathcal{A}^{(0)}( \lambda)- \mathcal{A}^{\delta}( \lambda)\big)(\mathcal{A}^{(0)})^{-1}(
\lambda)  \big]^2\partial_{\lambda} \mathcal{A}^{\delta}(
\lambda)\varphi~d\lambda+
$$
$$
\frac{1}{2i\pi}tr\int_{\partial{D_{\epsilon_{0}}}}
(\lambda-\lambda_{0}) \sum_{k\geq 3}(\mathcal{A}^{(0)})^{-1}(
\lambda)\big[
 \big(\mathcal{A}^{(0)}( \lambda)- \mathcal{A}^{\delta}( \lambda)\big)(\mathcal{A}^{(0)})^{-1}(
\lambda)  \big]^k\partial_{\lambda} \mathcal{A}^{\delta}(
\lambda)\varphi~d\lambda.
$$
Inserting expression (\ref{expan1}) into above relation, we may get:
\begin{equation}\label{lambda-lambda0}
\lambda_\delta-\lambda_0=
\frac{1}{2i\pi}tr\int_{\partial{D_{\epsilon_{0}}}}
(\lambda-\lambda_{0})(\mathcal{A}^{(0)})^{-1}\big[\delta\mathcal{A}^{(1)}(\mathcal{A}^{(0)})^{-1}+
\delta^2\mathcal{A}^{(2)}(\mathcal{A}^{(0)})^{-1}+\cdots\big]\big(\partial_{\lambda}
\mathcal{A}^{(0)}+\delta\partial_{\lambda}
\mathcal{A}^{(1)}\end{equation}$$+\delta^2\partial_{\lambda}
\mathcal{A}^{(2)}+\cdots\big)\varphi~d\lambda+
\frac{1}{2i\pi}tr\int_{\partial{D_{\epsilon_{0}}}}
(\lambda-\lambda_{0})(\mathcal{A}^{(0)})^{-1}\big[\delta\mathcal{A}^{(1)}(\mathcal{A}^{(0)})^{-1}+$$
$$
\delta^2\mathcal{A}^{(2)}(\mathcal{A}^{(0)})^{-1}+\cdots\big]^2\big(\partial_{\lambda}
\mathcal{A}^{(0)}+\delta\partial_{\lambda}
\mathcal{A}^{(1)}+\delta^2\partial_{\lambda}
\mathcal{A}^{(2)}+\cdots\big)\varphi~d\lambda+\cdots
$$
If we collect the same powers of $\delta$, then we get the desired
results. \square\\

Define,
$$
 \mathcal{B}_\delta(\lambda)\varphi(x) =(\mathcal{W}(\lambda)\varphi)(\Psi_{\delta}^{-1})(\Psi_{\delta}(x)),$$where
  $\varphi\in L^2(\partial\Omega)$ and $\Psi_\delta$ given by Section \ref{prob-form}, and $\mathcal{W}(\lambda)$
   is the operator associated to hydrodynamic double layer potential \cite{kohr,ladyzh,warn}. Then, the following main results hold.
\begin{theorem} \label{asym-v}
Let $\mathcal{A}_{\delta}$ be the operator defined by (\ref{op1}),
and
$\mathcal{M}_{\delta}=\mathcal{A}_{\delta}+\mathcal{B}_{\delta}$.
Let $\mathcal{O}_{0}$  be a  bounded neighborhood of
$\overline{\Omega}$ in  $\mathbb{R}^{3}$. Then there exists a
constant $\delta_{1}> 0$ smaller than $\delta_{0}$
 such that the
 eigenfunction $v(\delta) $ corresponding  to  the
eigenvalue,  $\lambda(\delta)$, in $(H^{1}(\Omega_{\delta}))^3\cap
H(\Omega_\delta) $
 can be chosen  to depend holomorphically in  $(x,\delta) \in
 \mathcal{O}_{0}\times ]-\delta_{1},\delta_{1}[$. Moreover this
eigenfunction satisfies  the following asymptotic formulae
\begin{equation}\label{asympt-fp}
 v(x;\delta) = v_{0}(x) + \sum_{n \geq 1}v_{n}(x)\delta^{n},
\end{equation} where the function $v_{0}$ is the eigenfunction solution of
(\ref{system1}) associated to $\lambda_{0}$. The terms $v_{n}$ are
computed from the Taylor coefficients of the operator valued
function $\mathcal{M}_{\delta}$ and  of those of the function
$a(\delta)=(a_{ij}(\delta))_{1\leq i,j\leq3}$.
\end{theorem}
\proof From \cite{kohr, medkova} we deduce that there exists a
continuous function $\varphi(t,\delta)$, which is analytic in
$\mathbb{R}^2\times]-\delta_{0},\delta_{0}[ $ and such that
\begin{equation}\label{eq-thm1}v_\delta(x) = \mathcal{S}(\lambda_\delta)\varphi+\mathcal{W}(\lambda_\delta)\varphi ,\quad x\in \Omega
\end{equation}
solves the eigenvalue problem  (\ref{system2}). Moreover, the
function given by $$U(\delta)(x)= \mathcal{M}(
\lambda(\delta))\varphi(\Psi^{-1}, \delta) $$ satisfies the
eigenvalue problem (\ref{system2})  in  $\Omega_{\delta}$ with the
boundary conditions: $U(\delta)|_{\partial{\Omega_{\delta}}}=0 $.
Here, $\mathcal{M}(
\lambda(\delta))\varphi(\Psi^{-1}, \delta) =\mathcal{M}_{\delta}(\lambda)\varphi.$\\
Now, by (\ref{eq-thm1}), we deduce that $v(x;\delta)=U(\delta)(x)=
\mathcal{M}( \lambda(\delta))\varphi(\Psi^{-1}, \delta) $ is jointly
analytic with respect to $(x,\delta)$ in $\{
\|x-\Psi_{\delta=0}(y)\| \leq z_{0}\} \times
]-\delta_{0},\delta_{0}[$, where $z_{0}$ is a positive constant. The
function $v(x;\delta)$ is jointly analytic in the variables
$(x,\delta) \in \mathcal{O}_{0}\times
]-\delta_{0},\delta_{0}[$.\\
We shall now give the asymptotic expansion of the function
$v(x;\delta)$ when $\delta$ tends to 0. Integral equation
(\ref{single}) gives us \bean \label{eq25}v(\delta)(x) =
\int_{\partial\Omega}M(\lambda(\delta),|x-
\Psi_{\delta}(y)|)\varphi(y,\delta)|\nabla\Psi_{\delta}(y)|d\sigma(y),
\eean where $M$ is the kernel of the operator
$\mathcal{M}_{\delta}$. The perturbed eigenvalue
 $\lambda(\delta)$ is in a small neighborhood of $\lambda_{0}$
for  small values of $\delta$. Then we have the following Taylor
expansion \bean M(\lambda(\delta),|x-
\Psi_{\delta}(y)|)|\nabla\Psi_{\delta}(y)|
 = M(\lambda_{0},|x- \Psi(y)|)|\nabla\Psi(y)| +
\sum_{k \geq 1}\delta^{k} M_{k}(x,y),\nonumber\\ \eean which holds
uniformly in $x \in \overline{\mathcal{O}}_{0}  $ and  $y \in
\partial\Omega$. The analyticity of the function $\varphi(y,\delta)$ with respect to $\delta$ immediately gives \bean \varphi(t,\delta) =
\varphi_{0}(y)  + \sum_{k\geq1}\delta^{k}\varphi_{k}(y), \eean
uniformly in $y \in \partial\Omega $. Substituting the last two
asymptotics into $(\ref{eq25})$ we find \bean
 v(x;\delta) =  v(x;\delta=0) +
\sum_{k\geq1}\delta^{k}[\sum_{n=1}^{k}\int_{\partial\Omega}\varphi_{k-n}(y)
M_{n}(x,y)d\sigma(y)]. \eean
  \square\\

The next result provide us with the asymptotic expansion of the
eigenpressures.

\begin{corollary}Suppose that we have all hypothesis of Theorem
(\ref{asym-v}). Then the eigenpressures $p_\delta$ solution of
(\ref{system2}) have the following uniform asymptotic expansion:
\begin{equation}\label{asym-pression}
 p_\delta(x) = p_{0}(x) + \sum_{n \geq 1}p_{n}(x)\delta^{n},
\end{equation} where the function $p_{0}$ is the eigenpressure solution of
(\ref{system1}) associated to $\lambda_{0}$. The terms $p_{n}$ are
computed from the Taylor coefficients $\lambda_n$ and
$v_n=(v_n^1,v_n^2,v_n^3)$ as follows:
$$\displaystyle p_n(.;x_i;.)=\int\Delta
v_n^idx_i+\sum_{k=0}^{n}\lambda_{k}\int v_{n-k}^{i}dx_i,\quad
\mbox{where }i=1,2,3.
$$
\end{corollary}
\proof From system (\ref{system2}) we have
$$\displaystyle \nabla.p_\delta=\Delta v_\delta+\lambda_\delta
v_\delta$$. Hence, we can expand the function $p_\delta$ in powers
of $\delta$ as we have done for $\lambda_\delta$ and $v_\delta$.
Moreover, we have:
\begin{equation}\label{as1-pression}\displaystyle \partial_ip_\delta=\Delta
v_\delta^i+\lambda_\delta v_\delta^i,\quad \mbox{for
}i=1,2,3.\end{equation}

 To get the coefficients of the formula (\ref{asym-pression}) one can insert both asymptotic expansions
(\ref{asympt-vp}) and (\ref{asympt-fp}) into relation
(\ref{as1-pression}), and integrate with the convenable variable.
 \square



\begin{thebibliography}{10}
\bibitem{albert1} J.H. Albert, {\em Genericity of simple eigenvalues for elliptic pde's,} Proc. Amer. Math.
Soc., 48 (1975), 413--418.
\bibitem{albert2} J.H. Albert, {\em Topology of the nodal and critical points sets for eigenfunctions of elliptic
operators,} Ph. D. Thesis, M. I. T., Boston MA, 1971.
\bibitem{AT} H. Ammari, F. Triki, {\em Splitting of resonant and scattering frequencies under shape deformation.} J. Differ. Equations 202, No. 2,
 (2004) 231-255.
\bibitem{ashbaugh} M. S. Ashbaugh, {\em On universal inequalities for the low eigenvalues of the buckling
problem. In Partial differential equations and inverse problems,}
volume 362 of Contemp. Math., pages 13--31. Amer. Math. Soc.,
Providence, RI, 2004.

\bibitem{babuska} I. Babuska, J.E. Osborn, Eigenvalue Problems, in: P.G. Ciarlet, J.L.
Lions (Eds.), Handbook of Numerical Analysis, vol. II, {\em Finite
Element Method} (Part I), North-Holland, Amsterdam, 1991, pp.
641?787.
\bibitem{nicolas} N. Depauw, {\em Solutions des \'equations de
Navier Stokes incompressibles dans un domaine exterieur } Rev. Mat.
Iberoam. 17(2001) 21--68.
\bibitem{kato} T. Kato, {\em Perturbation Theory for Linear Operators,} Springer-Verlag, Berlin, Heidelberg,
1980.
\bibitem{james} James. P. Kelliher, {\em Eigenvalues of the Stokes operator versus the Dirichlet Laplacian in the
plane.} Pac. J. Math. 244, No. 1, (2010) 99--132.
\bibitem{khelifi} A. Khelifi, {\em Asymptotic property and convergence estimation for the eigenelements of the Laplace operator.}  Appl. Anal. 86, No. 10, (2007) 1249--1264.
\bibitem{kohr} M. Kohr, {\em The interior Neumann problem for the Stokes resolvent system
in a bounded domain in $\mathbb{R}^n$,} Arch. Mech., 59, 3, (2007),
283--304.
\bibitem{ladyzh} O. A. Ladyzhenskaya, {\em Mathematical theory of the viscous
incompressible} id. Fizmathiz, Moscow, 1961, p. 203.
\bibitem{medkova} D. Medkova and W. Varnhorn, {\em Boundary value problems for the Stokes equations
with jumps in open sets}, Applicable Analysis, 87: 7, (2008, )829--
849.
\bibitem{ortega1} J. H. Ortega, and E. Zuazua, {\em  Generic simplicity of the eigenvalues of the stokes system in two soace
dimensions,} Advances in Differential Equations. Volume 6, Number 8,
(2001), 987--1023.
\bibitem{jia} Jia. Shanghu, Xie. Hehu, Yin. Xiaobo and Gao. Shaoqin
{\em Approximation and eigenvalue extrapolation of Stokes eigenvalue
problem by nonconforming finite element methods}. Applications of
Mathematics, vol. 54 (2009), issue 1, pp. 1--15
\bibitem{temam1} R. Temam, {\em Navier-Stokes Equations, Theory and Numerical Analysis,} Elsevier Science
Publishers B. V., Amsterdam, 1984.
\bibitem{temam2} R. Temam, {\em Navier-Stokes equations. AMS Chelsea Publishing, Providence, RI,
2001. Theory and numerical analysis,} Reprint of the 1984 edition.
\bibitem{warn} W. Varnhorn, {\em The Stokes equations, Akademie Verlag}, Berlin 1994.
\end{thebibliography}
\end{document}